\documentclass[aps,preprint,epsfig,rotate]{revtex4}
\begin{document}
\title{On the products of bipolar harmonics.}

 \author{Alexei M. Frolov}
 \email[E--mail address: ]{afrolov@uwo.ca}

\affiliation{Department of Applied mathematics \\
 University of Western Ontario, London, Ontario N6H 5B7, Canada}

\author{David M. Wardlaw}
 \email[E--mail address: ]{dwardlaw@mun.ca}

\affiliation{Department of Chemistry, Memorial University of Newfoundland, St.John's, 
             Newfoundland and Labrador, A1C 5S7, Canada}

\date{\today}

\begin{abstract}

The products of two and three bipolar harmonics ${\cal Y}^{\ell_1 \ell_2}_{LM}({\bf r}_{31}, {\bf r}_{32})$ are represented as the
finite sums of powers of the three relative coordinates $r_{32}, r_{31}$ and $r_{21}$. The complete (angular+radial) integrals of 
the products of the two and three bipolar harmonics in the basis of exponential radial functions are expressed as finite sums of the 
auxiliary three-particle integrals $\Gamma_{n,k,l}(\alpha, \beta, \gamma)$. The formulas derived in this study can be used to 
accelerate highly accurate computations of rotationally excited (bound) states in arbitrary three-body systems. In particular, we have 
constructed compact (400-term) variational wave functions for the triplet and singlet $2P(L = 1)-$states in light two-electron atoms 
and ions. Highly accurate calculations (20 - 21 stable decimal digits in the total energy) of the triplet and singlet $2P(L = 1)-$states 
in the two-electron Li$^{+}$, Be$^{2+}$, B$^{3+}$ and C$^{4+}$ ions are performed for the first time. 
      
\end{abstract}

\maketitle
\newpage

The bipolar harmonics ${\cal Y}^{\ell_1 \ell_2}_{LM}({\bf r}_{31}, {\bf r}_{32})$ \cite{Var} are extensively used in various methods developed 
for highly accurate solutions of different three-body problems arising in atomic, molecular and nuclear physics. The functions, Eq.(\ref{eq2}), 
are often used to represent the `angular dependence' of two-electron wave functions. The effectiveness of bipolar harmonics as `angular' 
functions follows from their explicit form which reflects a number of transparent physical ideas. There are a number of successful 
generalizations of bipolar harmonics to four- and five-body systems, where the three- and four-polar harmonics arise. In this study, however, 
we restrict our analysis to three-body systems only. In general, the three-body bipolar harmonics are written in the form
\begin{eqnarray}
 {\cal Y}^{\ell_1 \ell_2}_{LM}({\bf x}, {\bf y}) = x^{\ell_1}_{31} y^{\ell_2}_{32}
 \sum_{m_1 m_2} C^{LM}_{\ell_1 m_1 \ell_2 m_2} Y_{\ell_1 m_1}({\bf n}_x) 
 Y_{\ell_2 m_2}({\bf n}_y) \label{eq1}
\end{eqnarray}
where $C^{LM}_{\ell_1 m_1 \ell_2 m_2}$ are the Clebsh-Gordan coefficients, while ${\bf x}$ and ${\bf y}$ are two fundamental vectors specifying 
the relative positions of the three particles. The vectors ${\bf n}_x = \frac{{\bf x}}{x}$ and ${\bf n}_y = \frac{{\bf y}}{y}$ are the unit-norm 
vectors used as arguments in the spherical harmonics. Let us designate three particles in our three-particle system by the numbers 1, 2 and 3. The 
`natural' choice of the two fundamental vectors ${\bf x}$ and ${\bf y}$ in Eq.(\ref{eq1}) for an arbitrary three-body system is ${\bf x} = {\bf r}_3 
- {\bf r}_1 = {\bf r}_{31}$ and ${\bf y} = {\bf r}_3 - {\bf r}_2 = {\bf r}_{32}$. In this case each bipolar harmonic, Eq.(\ref{eq1}), takes the form 
${\cal Y}^{\ell_1 \ell_2}_{LM}({\bf r}_{31},{\bf r}_{32})$ and contains the vectors ${\bf r}_{31}$ and ${\bf r}_{32}$. Here and everywhere below in 
this study we assume that these two vectors are truly independent, i.e. ${\bf r}_{31} \ne \lambda {\bf r}_{32}$, where $\lambda$ is a numerical 
constant. Finally, the explicit form of the bipolar harmonics takes the form
\begin{eqnarray}
 {\cal Y}^{\ell_1 \ell_2}_{LM}({\bf r}_{31}, {\bf r}_{32}) = r^{\ell_1}_{31} r^{\ell_2}_{32} \sum_{m_1 m_2} C^{LM}_{\ell_1 m_1 \ell_2 m_2} 
 Y_{\ell_1 m_1}({\bf n}_{31}) Y_{\ell_2 m_2}({\bf n}_{32}) = r^{\ell_1}_{31} r^{\ell_2}_{32} {\cal Y}^{\ell_1 \ell_2}_{LM}({\bf n}_{31}, {\bf n}_{32}) 
 \label{eq2}
\end{eqnarray}

Variational wave functions which include bipolar harmonics are used to approximate the actual wave functions of bound states with non-zero angular 
momentum $L$. In many papers the bipolar harmonics are called and considered as the `angular parts' of basis functions. It is assumed that the additional 
`radial' part of the total wave function depends upon the three radial coordinates $r_{32}, r_{31}$ and $r_{21}$ only, i.e., it does not contain any of 
the angular variables. It is clear that the bipolar harmonics with the same $L$ and $M$ values (or indexes) form the $(2 L + 1)-$dimensional representation 
of the rotation group $SO(3)$. The explicit form of the matrices which describe transformations of the bipolar harmonics during rotations can be found with 
the use of Eq.(\ref{eq2}) and formulas from \cite{Edm} and \cite{Rose}. It can be shown that each matrix element is the product of two Clebsh-Gordan 
coefficients and two Wigner's $D-$functions. The explicit expression for these matrix elements can be reduced to another `short' form, but below we will not 
need these formulas.  
   
In this communication we develop the new method to operate with the bipolar harmonics. 
Our main interest below is related to the products of the two and three bipolar 
harmonics and angular integrals of such products. First, note that bipolar harmonics 
with the same $LM$ indexes form the closed algebra, i.e. the product of two bipolar 
harmonics ${\cal Y}^{\ell_1 \ell_2}_{L_a M_a}({\bf r}_{31}, {\bf r}_{32})$ and 
${\cal Y}^{\ell_3 \ell_4}_{L_b M_b}({\bf r}_{31}, {\bf r}_{32})$ is always represented 
as the finite sum of bipolar harmonics with the different values of $L_c$ and $M_c$. 
This can be written in the following form 
\begin{eqnarray}
  {\cal Y}^{\ell_1 \ell_2}_{L_a M_a}({\bf r}_{31}, {\bf r}_{32}) {\cal Y}^{\ell_3 \ell_4}_{L_b M_b}({\bf r}_{31}, {\bf r}_{32}) &=& 
  \sum_{\ell_{a} \ell_{b}} f_{LM}(\ell_{1}, \ell_{2}, \ell_{3}, \ell_{4}, \ell_{5}, \ell_{6}; L_a, M_a, L_b, M_b, L_c, M_c) \label{eq3} \\
  & & {\cal Y}^{\ell_{5} \ell_{6}}_{L_c M_c}({\bf r}_{31}, {\bf r}_{32}) \nonumber 
\end{eqnarray} 
where $f_{LM}(\ell_{1}, \ell_{2}, \ell_{3}, \ell_{4}; \ell_{a}, \ell_{b})$ are the numerical coefficients which can be determined, e.g., by mulitplying the 
both sides of Eq.(\ref{eq3}) by the different bipolar harmonics and performing integration of the both parts of arising equation over all angular variables. 

Let us briefly discuss the integration over the angular variables in an arbitrary non-relativistic system of three particles. In general, to describe the 
non-relativistic quantum system of spinless particles one needs 9 (3 $\times$ 3) dynamical variables. Three of these nine variables describe the translations of 
the solid triangle of particles. The internal state of three-body system does not change during such translations. In actual cases these three (Galilean) 
translations can be separated by using an appropriate choice of internal coordinates. Formally, we can assume that after such a separation of translations one of 
the three particles, e.g. the third particle, will always be at rest. The remaining six coordinates are separated into two groups: (a) three coordinates which are 
rotationally invariant, i.e. they do not change during any rotation of the whole three-body systems, and (b) three coordinates which describe rotations of the 
whole three-body system. The scalar coordinates which do not change during any rotation and/or translation of the three-body system can be chosen as the three 
interparticle distances $r_{32}, r_{31}, r_{21}$.

The choice of the three truly independent `rotational' coordinates in the three-body system is slightly more complicated, since such coordinates must be related 
with the angular coordinates of the two vectors ${\bf r}_{31}$ and ${\bf r}_{32}$. Let $\theta_1, \phi_1$ and $\theta_2, \phi_2$ be the spherical coordinates of 
these two vectors. In these coordinates for an elementary volume $dV$ we can write 
\begin{eqnarray}
 dV = r^{2}_{31} dr_{31} sin\theta_{31} d\theta_{31} d\phi_{31} r^{2}_{32} dr_{32} sin\theta_{32} d\theta_{32} d\phi_{32}
\end{eqnarray}
These six coordinates $(r_{31}, \theta_{31}, \phi_{31}, r_{32}, \theta_{32}, \phi_{32})$ 
can be used to describe an arbitrary three-body system. However, as it was shown by 
Hylleraas in 1929 \cite{Hyl} it is better to choose three-body coordinates in a different way. 
In \cite{Hyl} three radial variables were chosen as scalar interparticle distances (or 
interparticle coordinates) $r_{ij} = \mid {\bf r}_i -  {\bf r}_j \mid = r_{ji}$, where $(ij)$ 
= (32), (31), (21). On the other hand, it is clear that three angular variables can be chosen 
as the three Euler's angles $\phi_{31}, \theta_{31}, \phi_{32}$. In reality, one finds a number 
of advantages in calculations, if such a  `natural' choice of angular variables is used. In 
particular, in all earlier papers (see, e.g., \cite{Drake} - \cite{Fro1} and references therein) 
three radial $r_{32}, r_{31}, r_{21}$ and three angular variables $\phi_{31}, \theta_{31}, 
\phi_{32}$ were chosen as described here. In these variables an elementary volume $dV$ takes 
the form    
\begin{eqnarray}
 dV = r_{32} r_{31} r_{21} dr_{32} dr_{31} dr_{21} sin\theta_{31} d\theta_{31} d\phi_{32}
 d\phi_{31}
\end{eqnarray}   

This part of our study can be concluded with the two following comments. First, the three radial $r_{32}, r_{31}, r_{21}$ and three angular variables $\phi_{31}, 
\theta_{31}, \phi_{32}$ are semi-separated from each other. This means that the angular integral of any function of these six variables integrated over three angular 
variables (or over Euler's angles) is a function of the three radial variables only. Furthermore, in many actual cases these functions of radial variables are written in 
a relatively simple, finite-term form. Second, the three radial variables $r_{32}, r_{31}, r_{21}$ are not independent of each other, since e.g., $r_{21} \le r_{32} + 
r_{31}$ and $r_{31} \ge \mid r_{32} - r_{21} \mid$. Such constraints substantially complicate analytical and numerical calculations of three-body integrals. Therefore, in 
actual cases it is better to use three truly independent perimetric coordinates $u_1, u_2, u_3$  \cite{Pek}, which are related with the relative coordinates by the 
following linear transformation
\begin{eqnarray}
 u_1 = \frac12 (r_{31} + r_{21} - r_{21}) \; \; \; , \; \; \; u_2 = \frac12 (r_{32} + r_{21} - 
 r_{31}) \; \; \; , \; \; \; u_3 = \frac12 (r_{32} + r_{31} - r_{21}) \nonumber
\end{eqnarray}
The inverse relation takes the form
\begin{eqnarray}
r_{32} = u_2 + u_3  \; \; \; , \; \; \; r_{31} = u_1 + u_3 \; \; \; , \; \; \; r_{32} = u_1 + u_2 
\nonumber
\end{eqnarray}   
The Jacobian of the $(r_{32},r_{31},r_{21}) \rightarrow (u_1,u_2,u_3)$ transformation equals 2. The three perimetric coordinates $u_1, u_2, u_3$ are independent of each 
other and each of them varies between 0 and $+\infty$. 

The formula for the angular integral of the product of the two bipolar harmonics can be written in the form 
\begin{eqnarray}
 \oint d\Omega {\cal Y}^{\ell_1 \ell_2}_{LM}({\bf r}_{31}, {\bf r}_{32})
  {\cal Y}^{\ell_3 \ell_4}_{LM}({\bf r}_{31}, {\bf r}_{32}) = 
 F^{L}_{\ell_1 \ell_2; \ell_3 \ell_4}(r_{32}, r_{31}, r_{21}) \label{eq4}
\end{eqnarray}
The explicit form of the radial $F^{L}_{\ell_1 \ell_2; \ell_3 \ell_4}$ function can be found with the use of Eq.(\ref{eq2}) \cite{Drake}, \cite{Efr}. The result is
\begin{eqnarray}
 F^{L}_{\ell_1 \ell_2; \ell_3 \ell_4}(r_{32}, r_{31}, r_{21}) &=& 
 \frac12 (-1)^{L} r^{\ell_1 + \ell_3}_{31} r^{\ell_2 + \ell_4}_{32} 
 \sqrt{[\ell_1] [\ell_2] [\ell_3] [\ell_4]} \sum_{\lambda} (-1)^{\lambda} [\lambda]
 \left( \begin{array}{ccc} 
             \ell_1 & \ell_3 & \lambda \\ 0 & 0 & 0 \\
  \end{array} \right) \times \nonumber \\
 & & \left( \begin{array}{ccc} 
             \ell_2 & \ell_4 & \lambda \\ 0 & 0 & 0 \\
  \end{array} \right)
  \left\{ \begin{array}{ccc} 
             \ell_3 & \ell_4 & L \\ \ell_2 & \ell_1 & \lambda \\
  \end{array} \right\}
 P_{\lambda}(x)  \label{eq5}
\end{eqnarray}
where $[a] = 2 a + 1$ and the notation $P_{\lambda}(x)$ stands for the Legendre polynomial of the order $\lambda$, where $\lambda$ is a positive integer. Also, in this formula (and 
in some formulas below) we use the standard notations for the $3j-$ and $6j-$symbols \cite{LLQ}. The sum over $\lambda$ in Eq.(\ref{eq5}) is always finite, since the product of two 
$3j-$symbols is not zero only for those $\lambda$ which are bounded between the following values: $max \{\mid \ell_1 - \ell_3 \mid, \mid \ell_2 - \ell_4 \mid\} \le \lambda \le 
min\{\ell_1 + \ell_3, \ell_2 + \ell_4\}$. Moreover, the product of these two $3j-$symbols equals zero unless the two sums of the corresponding momenta ($\ell_1 + \ell_3 + \lambda$ 
and $\ell_2 + \ell_4 + \lambda$) are even numbers.   

The variable $x$, in Eq.(\ref{eq5}), is the following dimensionless ratio
\begin{eqnarray}
 x = \frac{r^2_{31} + r^2_{32} - r^2_{21}}{2 r_{31} r_{32}} \label{eq8}
\end{eqnarray}
This expression can be transformed with the use of the formula (8.911) from \cite{GR} for the Legendre polynomial $P_{\lambda}(x)$
\begin{eqnarray}
 P_{\lambda}(x) = \frac{1}{2^{\lambda}} \sum^{\Lambda}_{k=0} \frac{(-1)^k (2 \lambda 
 -2 k)!}{k! (\lambda - k)! (\lambda - 2 k)!} x^{\lambda - 2 k} = \frac{1}{2^{\lambda}} 
 \sum^{\Lambda}_{k=0} a_{\lambda,k} x^{\lambda - 2 k} \label{eq7}
\end{eqnarray} 
where $\Lambda = \Bigl[ \frac{\lambda}{2} \Bigr]$ is the integer part of 
$\frac{\lambda}{2}$ and coefficients $a_{\lambda,k}$ are
\begin{eqnarray}
 a_{\lambda,k} = \frac{(-1)^k (2 \lambda - 2 k)!}{k! (\lambda - k)! (\lambda - 2 k)!} 
 \nonumber
\end{eqnarray}
Now, by using the formula, Eq.(\ref{eq8}), one finds the following 
expression for the $x^{\lambda - 2 k}$ factor from Eq.(\ref{eq7}) 
\begin{eqnarray}
 x^{\lambda - 2 k} = r^{2 k - \lambda}_{31} r^{2 k - \lambda}_{32}
 \sum^{\lambda - 2 k}_{n=0} C^{n}_{\lambda - 2 k} (r^2_{32} - r^2_{21})^n
 r^{2\lambda - 4k - 2n}_{32} =  \sum^{\lambda - 2 k}_{n=0} C^{n}_{\lambda - 2 k} 
 \sum^{n}_{m=0} (-1)^m C^{m}_n \nonumber \\ 
 r^{\lambda - 2k - 2n}_{32} r^{2 k - \lambda + 2 n - 2 m}_{31} r^{2m}_{21} \label{eq9}
\end{eqnarray}
where $k \le \Lambda$ (see Eq.(\ref{eq7})) and notation $C^{a}_{b}$ stands for the binomial 
coefficients (the number of combinations from $b$ by $a$, where $a$ and $b$ are positive 
integer numbers).   

The formula, Eq.(\ref{eq9}), allows one to re-write the expression, Eq.(\ref{eq7}) in the form 
\begin{eqnarray}
 P_{\lambda}(x) = \frac{1}{2^{\lambda}} \sum^{\Lambda}_{k=0} a_{\lambda,k}  
 \sum^{\lambda - 2 k}_{n=0} C^{n}_{\lambda - 2 k} \sum^{n}_{m=0} (-1)^m C^{m}_n
 r^{\lambda - 2k - 2n}_{32} r^{2 k - \lambda + 2 n - 2 m}_{31} r^{2m}_{21} \label{eq10}
\end{eqnarray}
Now, by using the formulas, Eqs.(\ref{eq9}) and (\ref{eq10}), we can derive the following 
finite-sum expression for the exponential integral of the Legendre polynomial $P_{\lambda}(x)$
\begin{eqnarray}
 & &I_{\ell_2 + \ell_4; \ell_1 + \ell_3}(\alpha, \beta, \gamma; P_{\lambda}) = \int \int \int 
 P_{\lambda}(x) \exp(-\alpha r_{32} - \beta r_{31} - \gamma r_{21}) r^{\ell_2+\ell_4+1}_{32} 
 r^{\ell_1+\ell_3+1}_{31} r_{21} dr_{32} dr_{31} dr_{21} \nonumber \\
 &=& \frac{1}{2^{\lambda}} \sum^{\Lambda}_{k=0} a_{\lambda,k} \sum^{\lambda - 2 k}_{n=0} 
 C^{n}_{\lambda - 2 k} \sum^{n}_{m=0} (-1)^m C^{m}_n \Gamma_{\ell_2 + \ell_4 + \lambda - 2k 
 - 2n + 1, \ell_1 + \ell_3 + 2 k - \lambda + 2 n - 2 m + 1, 2 m + 1}(\alpha, \beta, \gamma) 
 \label{eq11}
\end{eqnarray}
where $\Gamma_{k,l,n}(a,b,c)$ is the basic three-body integral defined in \cite{Fro2012}. The definition of the basic three-body integral is written in the form
\begin{eqnarray}
 \Gamma_{k;l;n}(\alpha, \beta, \gamma) = \int \int \int r^{k}_{32} r^{l}_{31} 
 r^{n}_{21} \exp(-\alpha r_{32} - \beta r_{31} - \gamma r_{21})
 dr_{32} dr_{31} dr_{21} \label{e10}
\end{eqnarray}
where all indexes $k, l, n$ are assumed to be non-negative integer numbers. The analytical
formula used for numerical computations of such integrals is obtained from Eq.(\ref{e10})
by performing integration in perimetric coordinates \cite{Fro1}
\begin{eqnarray}
 & &\Gamma_{k;l;n}(\alpha, \beta, \gamma) = 2 \sum^{k}_{k_1=0} \sum^{l}_{l_1=0} 
 \sum^{n}_{n_1=0} C^{k}_{k_1} C^{l}_{l_1} C^{n}_{n_1} 
 \frac{(l-l_1+k_1)!}{(\alpha + \beta)^{l-l_1+k_1+1}}
 \frac{(k-k_1+n_1)!}{(\alpha + \gamma)^{k-k_1+n_1+1}}
 \frac{(n-n_1+l_1)!}{(\beta + \gamma)^{n-n_1+l_1+1}} \nonumber \\
 &=& 2 \cdot k! \cdot l! \cdot n! \sum^{k}_{k_1=0} \sum^{l}_{l_1=0} \sum^{n}_{n_1=0} 
 \frac{C^{k_1}_{n-n_1+k_1} C^{l_1}_{k-k_1+l_1} C^{n_1}_{l-l_1+n_1}}{(\alpha + 
 \beta)^{l-l_1+k_1+1} (\alpha + \gamma)^{k-k_1+n_1+1} (\beta + \gamma)^{n-n_1+l_1+1}} 
 \label{eq12}
\end{eqnarray}
where $C^{m}_{M}$ are the binomial coefficients. The formula, Eq.(\ref{eq12}), can also be written in a few other equivalent forms. The function $\frac{n!}{X^{n+1}}$ 
in Eq.(\ref{eq12}) is the $A_n(X)$ function introduced by Larson \cite{Lars}. The formula, Eq.(\ref{eq12}), was produced forthe first time by one of the author (AMF) 
in the middle of 1980's (see, e.g., \cite{Fro1} and references therein). The formula, Eq.(\ref{eq12}), has been used in calculations of various three-body integrals, 
e.g., integrals containing one or two Bessel functions \cite{Fro2012}. 

With the use of the formulas derived above one can obtain the closed (i.e. finite term) analytical formula for the following exponential integral
\begin{eqnarray}
 & &{\cal F}^{L}_{\ell_1 \ell_2; \ell_3 \ell_4}(a, b, c) = \int \int \int F^{L}_{\ell_1 \ell_2; 
 \ell_3 \ell_4}(r_{32}, r_{31}, r_{21}) \exp(-a r_{32} -b r_{31} -c r_{21}) r_{32} r_{31} r_{21}
 dr_{32} dr_{31} dr_{21} \nonumber \\
 & & dr_{31} dr_{21} = \frac12 (-1)^{L} \sqrt{[\ell_1] [\ell_2] [\ell_3] [\ell_4]} \sum_{\lambda} 
 (-1)^{\lambda} \frac{[\lambda]}{2^{\lambda}} \left( \begin{array}{ccc} 
             \ell_1 & \ell_3 & \lambda \\ 0 & 0 & 0 \\
  \end{array} \right) 
 \left( \begin{array}{ccc} 
             \ell_2 & \ell_4 & \lambda \\ 0 & 0 & 0 \\
  \end{array} \right)
  \left\{ \begin{array}{ccc} 
             \ell_3 & \ell_4 & L \\ \ell_2 & \ell_1 & \lambda \\
  \end{array} \right\} \times \nonumber \\
 & & \sum^{\Lambda}_{k=0} a_{\lambda,k} \sum^{\lambda - 2 k}_{n=0} C^{n}_{\lambda - 2 k} 
 \sum^{n}_{m=0} (-1)^m C^{m}_n \Gamma_{\ell_2 + \ell_4 + \lambda - 2k - 2n + 1, \ell_1 + \ell_3 
 + 2 k - \lambda + 2 n - 2 m + 1, 2 m + 1}(a, b, c) \label{eq13}
\end{eqnarray}
The derivation of this formula was the main goal of our study. This formula is of great interest
for numerical calculations of matrix elements which are needed to determine the total energies 
of bound states in three-body systems with $L \ge 1$ and calculate various expectation values. 
Recently, we have developed a number of fast numerical approaches to calculate the auxiliary 
three-particle integrals $\Gamma_{n,k,l}(a, b, c)$. This allows one to accelerate substantially all 
numerical calculations of matrix elements (see below). 

Note that the matrix elements of the potential energy are written in the form of Eq.(\ref{eq12}) 
only in those cases when all interparticle interaction potentials are the scalar functions of 
interparticle distances $r_{32}, r_{31}$ and $r_{21}$. Such cases include Coulomb three-body systems 
and three-body systems in which the potential energy is written in the form $V_a(r_{32}) + V_b(r_{31})
+ V_c(r_{21})$. In more complex cases the interaction potential between each pair of particles can 
also be a function of angular coordinates. Analytical formulas for the matrix elements in such 
cases must include angular integrals of the products of three bipolar harmonics. Such integrals are
discussed below.  

Now, let us present the results of numerical calculations with the use of formulas derived above. 
These results are shown in Tables I and II. In Table I we demonstrate the results of numerical 
calculations of the $I_{\ell_2 + \ell_4; \ell_1 + \ell_3}(\alpha, \beta, \gamma; P_{\lambda})$ and
${\cal F}^{L}_{\ell_1 \ell_2; \ell_3 \ell_4}(a, b, c)$ integrals, Eq.(\ref{eq13}), determined for 
different numerical values of its arguments ($a, b, c$) and parameters ($L, (\ell_1, \ell_2), (\ell_3, 
\ell_4)$). The method of numerical computations of these (exponential) integrals is based on the 
formula, Eq.(\ref{eq13}). However, for actual calculations this formula has been modified to avoid 
numerical instabilities which arise during summation of large numbers of positive and negative terms. 
The first formula, Eq.(\ref{eq11}), was re-written in the form
\begin{eqnarray}
 & & I_{\ell_2 + \ell_4; \ell_1 + \ell_3}(\alpha, \beta, \gamma; P_{\lambda}) 
 = \frac{1}{2^{\lambda}} \sum^{\Lambda}_{k=0} \mid a_{\lambda,k} \mid \sum^{\lambda - 2 k}_{n=0} 
 C^{n}_{\lambda - 2 k} \sum^{n}_{m=0} (-1)^{k+m} C^{m}_n \nonumber \\
 & & \Gamma_{\ell_2 + \ell_4 + \lambda - 2k 
 - 2n + 1, \ell_1 + \ell_3 + 2 k - \lambda + 2 n - 2 m + 1, 2 m + 1}(\alpha, \beta, \gamma) 
 \label{eq11a}
\end{eqnarray}
where $\Lambda = \Bigl[ \frac{\lambda}{2} \Bigr]$, while the coefficients $\mid a_{\lambda,k} \mid$
are
\begin{eqnarray}
 \mid a_{\lambda,k} \mid = \frac{(2 \lambda - 2 k)!}{k! (\lambda - k)! (\lambda - 2 k)!} 
 \nonumber
\end{eqnarray}
The sums of the positive and negative terms in Eq.(\ref{eq11a}) must be calculated separately. At the 
second step we have used the following formula
\begin{eqnarray}
 & &{\cal F}^{L}_{\ell_1 \ell_2; \ell_3 \ell_4}(a, b, c) = \frac12 (-1)^{L} 
 \sqrt{[\ell_1] [\ell_2] [\ell_3] [\ell_4]} \sum_{\lambda} 
 (-1)^{\lambda} [\lambda] \left( \begin{array}{ccc} 
             \ell_1 & \ell_3 & \lambda \\ 0 & 0 & 0 \\
  \end{array} \right) 
 \left( \begin{array}{ccc} 
             \ell_2 & \ell_4 & \lambda \\ 0 & 0 & 0 \\
  \end{array} \right) \times \nonumber \\
 & & \left\{ \begin{array}{ccc} 
             \ell_3 & \ell_4 & L \\ \ell_2 & \ell_1 & \lambda \\
  \end{array} \right\} \cdot
 I_{\ell_2 + \ell_4; \ell_1 + \ell_3}(\alpha, \beta, \gamma; P_{\lambda}) \label{eq5a}
\end{eqnarray}
These two formulas are used for very fast and accurate calculations of the exponential integrals 
${\cal F}^{L}_{\ell_1 \ell_2; \ell_3 \ell_4}(a, b, c)$ which can be found in each matrix element 
of the Hamiltonian and overlap matrices for the bound states with $L \ge 1$. By performing extensive 
numerical computations we have found that the method based on the modified formulas, Eqs.(\ref{eq11a}) 
- (\ref{eq5a}), is fast, numerically reliable and can be applied in computations of different 
rotationally excited states, including highly excited states with $L \ge 15 - 20$. 

In Table II we determine the total energies of the bound $P(L = 1)-$states of a number of three-body 
systems. These systems include the two-electron ${}^{\infty}$He, ${}^4$He and ${}^3$He atoms and
two-electron He-like ions: Li$^{+}$, Be$^{2+}$, B$^{3+}$ and C$^{4+}$. For the bound $P(L = 1)-$states 
we can perform highly accurate numerical calculations by using our old approach \cite{Fro1} and the new 
method described in this study. Therefore, we can compare the final accuracy of both methods and 
computational times needed to compute the same values.  The nuclear masses of the ${}^3$He and ${}^4$He 
nuclei are 5495.8852 $m_e$ and 7294.2996 $m_e$, respectively \cite{CRC}. The masses of the nuclei in 
all two-electron ions and the He atom are assumed to be infinite. For each of these systems 
we determine the total energies $E$ (in atomic units) of the singlet $2^1P-$states and triplet 
$2^3P-$states (see Table II). Our trial wave functions contain $N$ = 400 exponential basis functions. 
The explicit form of such wave functions is
\begin{eqnarray}
 \psi(r_{32}, r_{31}, r_{21}) = \frac{1}{\sqrt{2}} [1 + (-1)^{\epsilon} \hat{P}_{12}] \cdot 
 {\cal Y}^{\ell_1 \ell_2}_{10}({\bf r}_{31}, {\bf r}_{32}) \sum^{N}_{i=1} C_i exp(-\alpha_i r_{32} 
 - \beta_i r_{31} - \gamma_i r_{21}) \label{eqt7}
\end{eqnarray} 
where $C_i$ are the linear variational coefficients, $\epsilon = 1$ in the case of the triplet states
and $\epsilon = 2$ (or 0) in the case of the singlet states. The operator $\hat{P}_{12}$ is the 
permutation of the two identical particles (electrons 1 and 2) and $N$ is the total number of terms in 
the trial function. Analogous wave functions with $N$ = 700 basis functions will later be used as  
short-term cluster functions in our highly accurate computations of the bound $2^1P-$ and $2^3P-$states 
in these atomic systems. All such calculations are usually performed with the use of our two-stage 
optimization strategy \cite{Fro01}. 

Preliminary results of highly accurate computations of the bound $2^1P-$ and $2^3P-$states in the Li$^{+}$, 
Be$^{2+}$, B$^{3+}$ and C$^{4+}$ ions are shown in Table III. In these calculations we have used the 
short-term cluster wave functions with $N = 400$ terms from Table II. The total number of basis functions
used in our highly accurate computations was varied between 2500 and 2850 exponential functions. More 
accurate calculations of these states are possible at this moment, but they require larger computational 
resources than currently availiable to the authors. The total energies and other bound state properties of the 
bound  $2^1P-$ and $2^3P-$states in the Li$^{+}$, Be$^{2+}$, B$^{3+}$ and C$^{4+}$ ions have never been determined 
to high accuracy (these bound states play important roles in some applications). The results from Table III are 
preliminary, but they will be used to accelerate the following highly accurate computations of the bound $P(L = 
1)-$states in these ions and other three-body systems. Highly accurate results (total energies) for the singlet 
and triplet $P(L = 1)-$states in the He atom(s) can be found in \cite{Fro2011}. 

As follows from the results of our highly accurate computations of the singlet and triplet $2P(L = 1)-$states in a 
number of two-electron ions we obtain a level of accuracy with the approach developed here that is very close to 
the accuracy of an earlier method from \cite{Fro1} which was specifically oriented to calculate the bound states
in three-body systems with small angular momenta $L$. Computational times for both methods are also 
comparable to each other. This indicates a very high efficiency of our current approach for bound states
with small $L$. Plus, now we have analytical formulas for bound three-body states with arbitrary $L$. Our 
computational interest in the bound $P(L = 1)-$states is based on the following facts. First, only for the
bound $P(L = 1)-$states can one find results determined to very high numerical accuracy, 
comparable to the accuracy known for the ground and low-lying excited $S(L = 0)-$states. Second, the 
problem of optimization of the non-linear parameters in Eq.(\ref{eqt7}) can be solved in a very fast and
accurate way for all bound states with $L = 1$. Formally, for bound $P-$states there is no difference
in our optimization algorithms from the case of the ground state(s) in three-body systems. For rotationally 
exicted states with $L \ge 2$ the process of optimization of the non-linear parameters is
significantly more complicated and takes substantial computational times. For instance, for the bound
$3D-$states in the He atom our current method produces the following total energies: -2.055620 7328528(4) 
$a.u$. (singlet) and -2.055636 3094537(4) $a.u$. Note that such an accuracy for these states is not
very high (it is comparable to the accuracy known for these states at the end of 1990's). To obtain 
better overall accuracy one needs to use a better optimization technique for the non-linear parameters in the
trial wave functions. On the other hand, all highly accurate calculations of the bound states with $L \ge 2$ 
can be performed with the use of the quadruple precision only. Even in calculations of the bound $D-$states 
with $N$ = 3500 - 4000 ($N$ is the total number of basis functions) there is no need to use the extended 
arithmetical precision \cite{Bail1}, since the coresponding overlap matrixes are not ill-conditioned. On the
other hand, the total energies and other bound state properties of such states are determined (with these 
wave functions) to the accuracy 14 - 15 decimal digits. Briefly, we can say that the new optimization and 
computational strategies must be developed for the rotationally excited bound states with $L \ge 2$ to 
produce results which contain 20 - 25 stable decimal digits. Right now, we do not have such strategies and 
this is the main reason why we have restricted this study to the bound $P-$states only.

Note that the matrix elements of the potential energy are written in the form of Eq.(\ref{eq12}) 
only in those cases when all interparticle interaction potentials are scalar functions of 
interparticle distances $r_{32}, r_{31}$ and $r_{21}$. Such cases include Coulomb three-body systems 
and three-body systems in which the potential energy is written in the form $V_a(r_{32}) + V_b(r_{31})
+ V_c(r_{21})$. In more complex cases the interaction potential between each pair of particles can 
also be a function of angular coordinates. In the general case, such potentials can be approximated by 
the sums of bipolar harmonics with the different $L$ and $M$ values. Analytical formulas for the matrix 
elements in such cases must include angular integrals of the products of three bipolar harmonics. 

Let us discuss the formulas for the products of three bipolar harmonics ${\cal Y}^{\ell_1 
\ell_2}_{L_aM_a}({\bf r}_{31}, {\bf r}_{32}), {\cal Y}^{\ell_3 \ell_4}_{L_bM_b}({\bf r}_{31}, {\bf 
r}_{32})$ and ${\cal Y}^{\ell_5 \ell_6}_{L_cM_c}({\bf r}_{31}, {\bf r}_{32})$. The general formulas
for such products can be found in \cite{Var}. It is clear that the angular integral of the product 
of three bipolar harmonics must be proportional to the Clebsh-Gordan coefficient $C^{L_aM_a}_{L_bM_b 
L_cM_c}$, or to the corresponding $3jm-$symbol (see below). Second, as we have mentioned above, the 
$(2 L + 1)$ bipolar harmonics ${\cal Y}^{\ell_1 \ell_2}_{LM}({\bf r}_{31}, {\bf r}_{32})$ (with 
the same $L$, but different $M$) are the basis vectors of the $(2 L + 1)$-dimensional representation 
of the rotation group. Therefore, as it follows from Schur's lemma the angular integral of the product 
of two bipolar harmonics ${\cal Y}^{\ell_1 \ell_2}_{L_aM_a}$ and ${\cal Y}^{\ell_1 \ell_2}_{L_bM_b}$ is 
always zero unless the values of $L_a, M_a$ and $L_b, M_b$ are exactly the same, i.e. $L_a = L_b$ and 
$M_a = M_b$. This explains the explicit form of the angular integral used in Eq.(\ref{eq4}). For the 
product of three bipolar harmonics the situation is different and we cannot assume $a$ $priori$ that 
even some of the indexes are equal. In general, the angular integral of the product of three bipolar 
harmonics is written in the form  
\begin{eqnarray}
 && \oint d\Omega {\cal Y}^{\ell_1 \ell_2}_{L_aM_a}({\bf r}_{31}, {\bf r}_{32})
  {\cal Y}^{\ell_3 \ell_4}_{L_bM_b}({\bf r}_{31}, {\bf r}_{32}) 
  {\cal Y}^{\ell_5 \ell_6}_{L_cM_c}({\bf r}_{31}, {\bf r}_{32}) 
  = \left( \begin{array}{ccc} 
             L_a & L_b & L_c \\ M_a & M_b & M_c \\
  \end{array} \right) \times \nonumber \\
 && G^{L_{a} L_{b} L_{c}}_{\ell_2,\ell_4,\ell_6; \ell_1,\ell_3,\ell_5}(r_{32}, r_{31}, r_{21}) =
  \left( \begin{array}{ccc} 
           L_a & L_b & L_c \\ M_a & M_b & M_c \\
  \end{array} \right) \sum_{\lambda} b_{\lambda} r^{\ell_2 + \ell_4 + \ell_6}_{32}
  r^{\ell_1 + \ell_3 + \ell_3}_{31} P_{\lambda}(x) \label{eq14}
\end{eqnarray}
where the function $G^{L_{a} L_{b} L_{c}}_{\ell_2,\ell_4,\ell_6;\ell_1,\ell_3,\ell_5}(r_{32}, r_{31}, 
r_{21})$ depends upon three relative coordinates $r_{32}, r_{31}$ and $r_{21}$. The explicit formula 
for this function is obtained from the last equality in Eq.(\ref{eq14}). The coefficient $b_{\lambda}$ 
in Eq.(\ref{eq14}) does not depend upon the relative coordinates, but it is a functions of all ten values 
of angular momenta $\lambda, L_{a}, L_{b}, L_{c}, \ell_1, \ell_2, \ell_3, \ell_4, \ell_5, \ell_6$. The 
formula for these coefficients takes the form
\begin{eqnarray}
 b_{\lambda} &=& \frac{1}{8 \pi} (-1)^{L_b + \lambda} \sqrt{[L_a] [L_b] [L_c] 
 [\ell_1] [\ell_2] [\ell_3] [\ell_4] [\ell_5] [\ell_6]} \cdot [\lambda] 
 \sum_{\lambda_1} \sum_{\lambda_2} 
  \left( \begin{array}{ccc} 
             \ell_1 & \ell_5 & \lambda_1 \\ 0 & 0 & 0 \\
  \end{array} \right) 
 \left( \begin{array}{ccc} 
             \ell_2 & \ell_6 & \lambda_2 \\ 0 & 0 & 0 \\
  \end{array} \right) \times \nonumber \\ 
 & & \left( \begin{array}{ccc} 
             \lambda & \ell_3 & \lambda_1 \\ 0 & 0 & 0 \\
 \end{array} \right)
 \left( \begin{array}{ccc} 
             \lambda & \ell_4 & \lambda_2 \\ 0 & 0 & 0 \\
 \end{array} \right) 
 \left\{ \begin{array}{ccc} 
             \ell_3 & \ell_4 & L_b \\ \lambda_2 & \lambda_1 & \lambda \\
  \end{array} \right\} 
 \left\{ \begin{array}{ccc} 
             \ell_3 & \ell_4 & L_a \\ 
             \ell_5 & \ell_6 & L_c \\ 
             \lambda_1 & \lambda_2 & L_b \\
  \end{array} \right\} \; \; \; .\label{eq15}
\end{eqnarray}
In this form Eqs.(\ref{eq14}) - (\ref{eq15}) look very similar to Eq.(\ref{eq4}). Numerical calculations
of the $b_{\lambda}$ coefficients with the use of Eq.(\ref{eq15}) is straightforward. For instance, for
$\ell_1 = 1, \ell_2 = 3, \ell_3 = 2, \ell_4 = 2, \ell_5 = 2, \ell_6 = 2$ and for $L_a = 1, L_b = 2, L_c = 1$ 
one finds from Eq.(\ref{eq15}) $b_{1} =  0.15921224404155089\cdot 10^{-1}, b_{3} = -0.67600638318413508 \cdot 
10^{-2}$ and $b_{5} = 0.27451117819426586 \cdot 10^{-2}$. The coefficients $b_{\lambda}$ with other values of 
$\lambda$ equal zero identically.

Finally, as it seen from the formulas, Eqs.(\ref{eq14}) - (\ref{eq15}) and Eq.(\ref{eq10}), the calculation 
of the three-body integralas which contain the products of three bipolar harmonics is reduced to the 
computation of some finite sums of the basic (or auxiliary) three-body integrals $\Gamma_{k;l;n}(\alpha, 
\beta, \gamma)$, Eq.(\ref{eq12}). The explicit expression for the radial integral of the 
$G^{L_{a} L_{b} L_{c}}_{\ell_2,\ell_4,\ell_6; \ell_1,\ell_3,\ell_5}(r_{32}, r_{31}, r_{21})$
function is 
\begin{eqnarray}
 T^{L_{a} L_{b} L_{c}}_{\ell_2,\ell_4,\ell_6; \ell_1,\ell_3,\ell_5} &=& 
 \int \int \int G^{L_{a} L_{b} L_{c}}_{\ell_2 + \ell_4 + \ell_6; \ell_1 + \ell_3 + \ell_5}(r_{32}, 
 r_{31}, r_{21}) r_{32} r_{31} r_{21} dr_{32} dr_{31} dr_{21} \nonumber \\
 &=& \sum_{\lambda} b_{\lambda} I_{\ell_2 + \ell_4 + \ell_6; \ell_1 + \ell_3 + \ell_5}(\alpha, \beta, 
 \gamma; P_{\lambda}) \; \; \; , \label{eq15a}
\end{eqnarray}
where $I_{\ell_2 + \ell_4 + \ell_6; \ell_1 + \ell_3 + \ell_5}(\alpha, \beta, \gamma; P_{\lambda})$ is defined by Eq.(\ref{eq11a}). This is another result 
which is of great interest for highly accurate computations of many actual three-body systems. Our test calcualtions of the $T^{L_{a} L_{b} 
L_{c}}_{\ell_2,\ell_4,\ell_6; \ell_1,\ell_3,\ell_5}$ coefficient performed for $\ell_1 = 1, \ell_2 = 3, \ell_3 = 2, \ell_4 = 2, \ell_5 = 2, \ell_6 = 2, 
L_a = 1, L_b = 2, L_c = 1$ lead to the following result: $T^{1 2 1}_{3,2,2;1,2,2} = 1.9488412125971230 \cdot 10^{4}$.

We have considered the products of the two and three bipolar harmonics ${\cal Y}^{\ell_1 \ell_2}_{LM}({\bf r}_{31}, {\bf r}_{32})$. It is shown that 
angular integrals of such products are represented as the finite sums of powers of the three relative coordinates $r_{32}, r_{31}$ and $r_{21}$ (or 
interpartilce distances). The six-dimensional (angular + radial) integrals of the products of the two and three bipolar harmonics in the basis of exponential 
radial functions are expressed as finite sums of the auxiliary three-particle integrals $\Gamma_{n,k,l}(\alpha, \beta, \gamma)$. The formulas derived in this 
study can be used to accelerate highly accurate computations of rotationally excited (bound) states in arbitrary three-body systems. The methods developed in 
this study have been used to construct very compact (but highly accurate!) variational wave functions of triplet and singlet $2P(L = 1)-$states in light 
two-electron atoms and ions. The preliminary results of our highly accurate calculations of the triplet and singlet $2P(L = 1)-$states in the two-electron 
Li$^{+}$, Be$^{2+}$, B$^{3+}$ and C$^{4+}$ ions contain 20 - 21 stable decimal digits. This makes our wave functions among the most accurate wave functions ever 
known for these atomic systems.

\newpage
\begin{table}[tbp]
   \caption{Numerical values of the $I_{\ell_2 + \ell_4; \ell_1 + \ell_3}(a, b, c; P_{\lambda})$
            and ${\cal F}^{L}_{\ell_1 \ell_2; \ell_3 \ell_4}(a, b, c)$ integrals computed
            for different values of the $\ell_1, \ell_2, \ell_3, \ell_4, \lambda$ and $L$
            parameters. $a = 1.55, b = 1.33$ and $c = 1.07$ in all cases.}
     \begin{center}
     \begin{tabular}{| c | c | c | c | c | c | c | c | c | c | c |}
      \hline\hline
 $\ell_1$ & $\ell_2$ & $\ell_3$ & $\ell_4$ & $\lambda$ & &  
        $I_{\ell_1 \ell_2; \ell_3 \ell_4}(a, b, c; P_{\lambda})$ & &
        $L$ & & ${\cal F}^{L}_{\ell_1 \ell_2; \ell_3 \ell_4}(a, b, c)$ \\
          \hline     
  1 & 0 & 1 & 0 & 1 & & 0.20963264469930568E+00 & & 1 & & 0.13008789710134043E+00 \\ 

  1 & 1 & 1 & 1 & 1 & & 0.11032101479039074E+01 & & 2 & & 0.72370337608449528E+00 \\ 

  1 & 1 & 2 & 0 & 1 & & 0.11491009319449536E+01 & & 2 & & 0.69783486804905589E+00 \\ 

  1 & 2 & 2 & 1 & 2 & & 0.33053241063178740E+02 & & 3 & & 0.83514739108659169E+01 \\ 

  1 & 2 & 1 & 2 & 2 & & 0.34640478424614411E+02 & & 3 & & 0.69505739570548458E+01 \\ 

  3 & 0 & 1 & 2 & 2 & & 0.18295480775566101E+02 & & 3 & & 0.14561365717576028E+02 \\ 

  3 & 0 & 0 & 3 & 3 & & 0.58890134034481354E+02 & & 3 & & 0.29445067017240677E+02 \\ 
    \hline\hline
  \end{tabular}
  \end{center}
  \end{table}
\begin{table}[tbp]
   \caption{The total energies $E$ of the bound ${}^1P(L = 1)-$ and ${}^3P(L = 1)-$states 
            of some two-electron atoms and ions (in atomic units). The total number 
            of basis functions used to construct these short-term wave functions is 400.}
     \begin{center}
     \begin{tabular}{| c | c | c | c |}
      \hline\hline
     & $E({}^1P(L = 1)-$state) & & $E({}^3P(L = 1)-$state) \\
          \hline     
 ${}^{\infty}$He & -2.12384 308649 749 & & -2.13316 419077 725 \\
      \hline
 ${}^{4}$He & -2.12354 565412 918 & & -2.13288 064210 349 \\
     
 ${}^{3}$He & -2.12344 834501 190 & & -2.13278 787470 796 \\
       \hline
 ${}^{\infty}$Li$^{+}$  & -4.99335 107777 845 & & -5.02771 568139 695 \\
       
 ${}^{\infty}$Be$^{2+}$ & -9.11077 162291 325 & & -9.17497 314304 428 \\
      
 ${}^{\infty}$B$^{3+}$  & -14.4772 832652 859 & & -14.5731 376921 778 \\
      
 ${}^{\infty}$C$^{4+}$  & -21.0933 323133 828 & & -21.2217 106964 635 \\
    \hline\hline
  \end{tabular}
  \end{center}
  \end{table}
\newpage
\begin{table}[tbp]
   \caption{Highly accurate total energies $E$ of the bound ${}^1P(L = 1)-$ and 
            ${}^3P(L = 1)-$states of some two-electron ions (in atomic units). 
            The total number of basis function is designated by $N$.}
     \begin{center}
     \begin{tabular}{| c | c | c | c | c |}
      \hline\hline
   ion     &  $N$  & $E({}^1P(L = 1)-$state)   &   $E({}^3P(L = 1)-$state)  \\
          \hline     
 Li$^{+}$   & 2500  &  -4.99335 10777 80017 36235 & -5.02771 56813 97367 762165 \\  

 Li$^{+}$   & 2700  &  -4.99335 10777 80017 36242 & -5.02771 56813 97367 762174 \\  

 Li$^{+}$   & 2850  &  -4.99335 10777 80017 36245 & -5.02771 56813 97367 762180 \\  
                         \hline\hline
 Be$^{2+}$  & 2500  &  -9.11077 16622 91644 408257 & -9.17497 31430 70973 000582 \\  

 Be$^{2+}$  & 2700  &  -9.11077 16622 91644 408262 & -9.17497 31430 70973 000594 \\  

 Be$^{2+}$  & 2850  &  -9.11077 16622 91644 408265 & -9.17497 31430 70973 000601 \\  
                         \hline\hline
 B$^{3+}$   & 2500  & -14.47728 32653 07799 28311 & -14.57313 76922 13480 04811 \\  

 B$^{3+}$   & 2700  & -14.47728 32653 07799 28988 & -14.57313 76922 13480 04813 \\  

 B$^{3+}$   & 2850  & -14.47728 32653 07799 29427 & -14.57313 76922 13480 04814 \\  
                         \hline\hline
 C$^{4+}$   & 2500  & -21.09333 23133 88409 05480 & -21.22171 06964 88051 07794 \\  

 C$^{4+}$   & 2700  & -21.09333 23133 88409 05491 & -21.22171 06964 88051 07766 \\ 

 C$^{4+}$   & 2850  & -21.09333 23133 88409 05510 & -21.22171 06964 88051 07771 \\ 
    \hline\hline
  \end{tabular}
  \end{center}
  \end{table}
\end{document}